\documentstyle[psfig,gcdv]{article}
\def\reference{\parskip 0pt\par\noindent\hangindent 0.5 truecm}

\def\spose#1{\hbox to 0pt{#1\hss}}
\def\simlt{\mathrel{\spose{\lower 3pt\hbox{$\mathchar"218$}}
     \raise 2.0pt\hbox{$\mathchar"13C$}}}
\def\simgt{\mathrel{\spose{\lower 3pt\hbox{$\mathchar"218$}}
     \raise 2.0pt\hbox{$\mathchar"13E$}}}

\newcommand{\sun}{\odot}
\newcommand{\be}{\begin{equation}}
\newcommand{\ee}{\end{equation}}
\newcommand{\etal}{et~al.\ }
\newcommand{\msun}{\mbox{M$_{\sun}$}}

\newcommand{\natd}[2]{\mbox{$#1 \cdot 10^{#2}$}}
\newcommand{\pder}[2]{\frac{\partial #1}{\partial #2}}
\newcommand{\pdert}[1]{\pder{#1}{t}}

\begin{document}

\small
\shorttitle{Can dust destabilize galactic disks?}
\shortauthor{Ch.\ Theis \& N. Orlova}
%
%
\title{\large \bf
Can dust destabilize galactic disks?}

\author{\small 
 Christian Theis$^{1}$,
 Natalya Orlova$^{2}$
} 

\date{}
\twocolumn[
\maketitle
\vspace{-20pt}
\small
{\center
  $^1$ Institut f\"ur Theoretische Physik und Astrophysik d.\ Univ.\ Kiel,
    D--24098 Kiel, Germany (theis@astrophysik.uni-kiel.de)\\
  $^2$ Institute of Physics, Stachki 194, Rostov-on-Don, Russia
     (orlova@rsusu1.rnd.runnet.ru)\\[3mm]
}

%
\begin{center}
{\bfseries Abstract}
\end{center}
\begin{quotation}
\begin{small}
\vspace{-5pt}
%
We studied the dynamical influence of a dust component on the gaseous phase in 
central regions of galactic disks. Therefore, we performed two-dimensional 
hydrodynamical simulations for flat multi-component disks embedded in a stellar 
and dark matter potential. The pressure-free dust component 
is coupled to the gas by a drag force depending on their velocity difference.

   It turned out that the most unstable regions are those with either a low 
or near to minimum Toomre parameter or with rigid rotation, i.e.\ the central 
area. In that regions the dust-free disks become most unstable for a small range
of high azimuthal modes ($m \sim 8$), whereas in dusty disks all modes have 
similar amplitudes resulting in a patchy appearance. The structures in 
the dust have a larger contrast between arm and inter-arm regions than
those of the gas. The dust peaks are frequently correlated
with peaks of the gas distribution, but they do not necessarily coincide 
with them. This leads to a large scatter in the dust-to-gas ratios.
The appearance of the dust is more cellular (i.e.\ sometimes connecting
different spiral features), whereas the gas is organized in a multi-armed
spiral structure.

   We found that an admixture of 2\% dust (relative to the mass of the gas)
destabilizes gaseous disks substantially, whereas dust-to-gas ratios below
1\% have no influence on the evolution of the gaseous disk. For a
high dust-to-gas ratio of 10\% the instabilities reach the saturation
level already after 30 Myr.


\end{small}
\end{quotation}
]

\bigskip

\begin{figure*}[hpt]
 \begin{center}
   \begin{tabular}{cc}
     \psfig{file=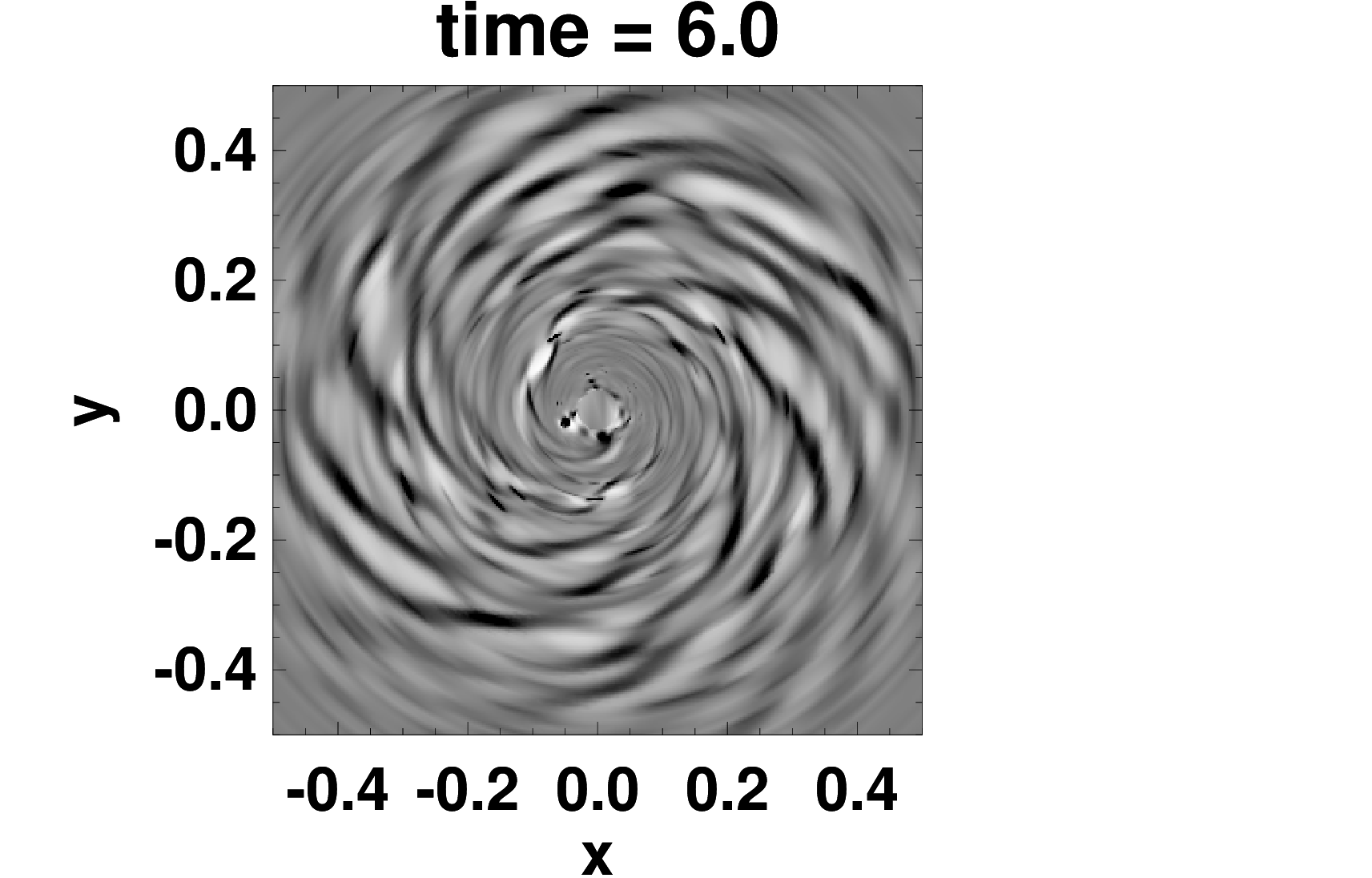,width=8.25cm,angle=90} &
     \psfig{file=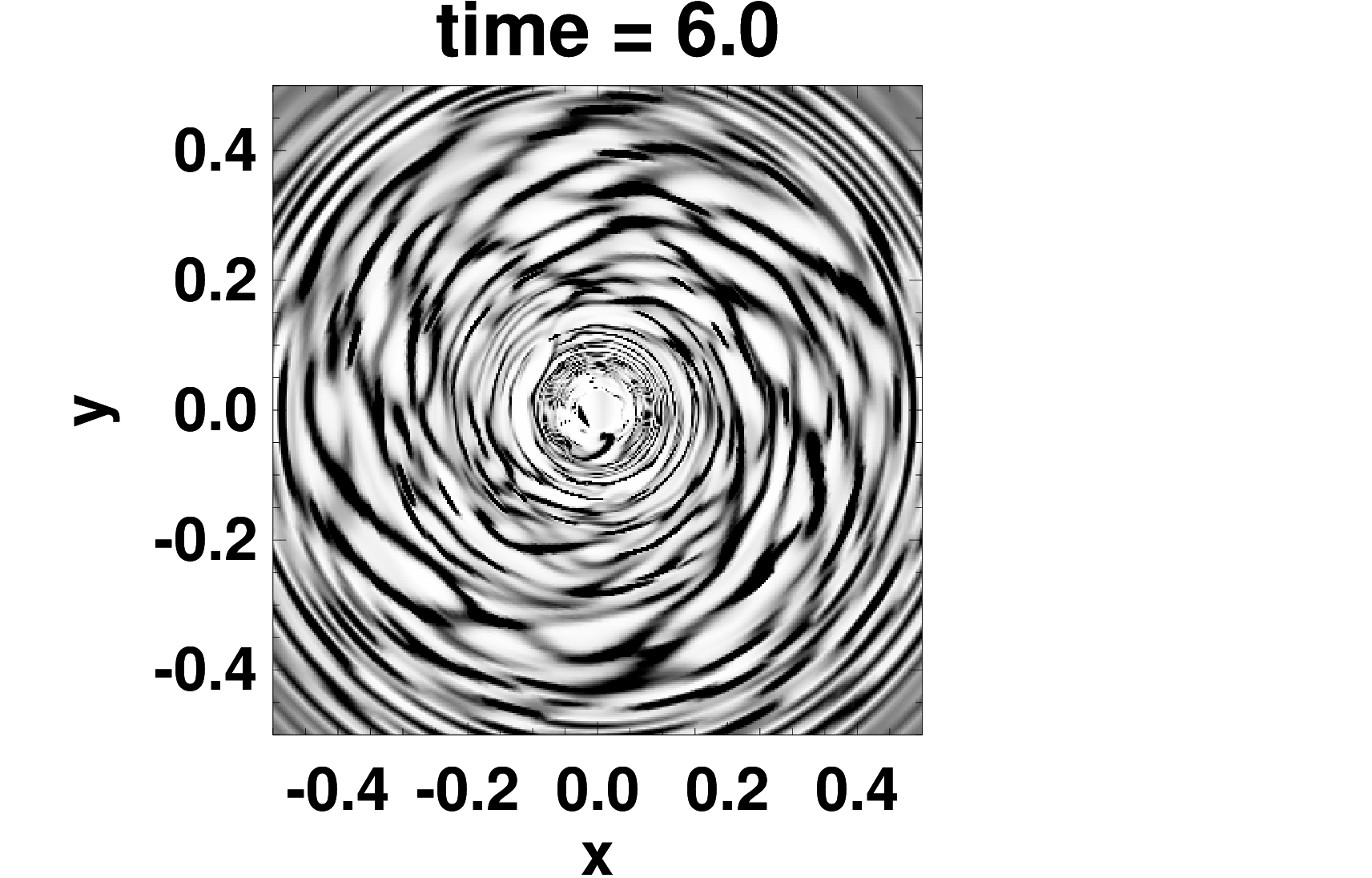,width=8.25cm,angle=90} \\
   \end{tabular}
   \caption{Perturbations of the surface density
    (normalized to their initial values) of the
    gas and dust component after $t= 6 \sim 90$ Myr for a
    model with a 10\% dust mass fraction (relative to gas):
    $\Delta \Sigma_{g,d} / \Sigma_{g,d}$ for gas (left) and 
    dust (right). Areas devoid of material are white, areas with a density 
    enhancement of a factor of 2 or more are black. The grey area seen at 
    the outer edges correspond to no deviation from the initial surface 
    density. The lengths at the $x-$ and $y-$axes are given in kpc.
   \label{muchdust_surface_denspert_image}}
 \end{center}
\end{figure*}

\section{Introduction}

  Chemodynamical simulations are characterized by a multi-phase treatment
of the interstellar medium (ISM). In most of these calculations the ISM 
is split into two dynamically different components, the clumpy 
star-forming molecular clouds and the diffuse warm or hot gas
(Theis et al.\ 1992, Samland et al.\ 1997). More refined decompositions
of the ISM are only done for one-zone models
(e.g.\ Ikeuchi et al.\ 1984), but not for gravitationally coupled,
dynamically evolving systems. Especially, the dust component in 
galaxies is usually just seen as a tracer population which has no dynamical
influence on galactic dynamics. On the other hand, it is well known
that even a small admixture of a dynamically cold component can substantially 
destabilize galactic structures as Jog \& Solomon (1984) demonstrated
for a small amount of cold gas embedded in stellar disks.
The aim of our paper is to study, whether a dynamically cold dust phase
can also destabilize galactic disks. 

  Good places for looking on the influence of dust are galactic central
regions. During the last years high resolution observations 
revealed that the circumnuclear regions contain stellar-gaseous mini-disks 
with sizes of a few hundred pc (e.g.\ Carollo et al.\ 1997).
These disks have an astonishing rich structure described as 
mini-bars, spiral-like dust lanes, star-forming rings or spiral arms. 
Many nuclear regions are also well described as patchy or multi-armed.
Investigations by Regan \& Mulchaey (1999) showed that these spirals are 
the most common morphological structures in the central regions of galaxies. 

 Since the nature of the mini-spirals is of great importance for our 
understanding of a variety of astrophysical processes like the mass 
accretion into galactic nuclei, several explanations for have been invoked. 
E.g.\ Athanassoula (1992) studied the gas flows in and around bars by 
(stellar-dynamical) orbital analysis. She found that the existence of 
different periodic orbit families (or equivalently the existence of inner 
Lindblad resonances (ILR)) is a key criterion for the existence and the
shape of dust lanes. 

Different to the stellar-dynamical interpretation Engl\-maier \& 
Shlosman (2000) suggested that mini-spirals in central regions of 
galactic disks
are related to the formation of grand-design spiral patterns in galaxies.
They argued that gas density waves -- different to stellar density waves --
are not completely damped or absorbed at the ILR and, thus, they may 
generate spiral structures at all radii including the nuclear regions. 
Such a model might explain the continuity of some spiral features at small
and large radii as well as the low arm-interarm contrast observed
in galactic centers. 
A similar idea of induced structure formation is to invoke secondary
bars or small nuclear bars located inside the ILR 
(Wada \& Koda 2001).

A common property of all the mentioned mechanisms is that they
result in structures dominated by two or a few arms. 
On the other hand, Elmegreen \etal (2002) stressed that nuclear dust spirals
differ from main-disk spirals in several respects: the nuclear spirals are 
very irregular with both trailing and leading components that often cross.

All discussed mechanisms have some weak points and can not fully 
account for all the observational data. We consider here a new approach 
based on the observation that circumnuclear disks of galaxies are dusty.
By means of a stability analysis Noh \etal (1991) showed
that a dust component can strongly destabilize proto-planetary disks. 
The admixture of only 2\% of dust enhances the growth rates
of the dominant gaseous phase significantly. A conservative estimate of the
dust-to-gas ratio $r$ in the solar neighborhood gives an average value of 
0.6\% ranging from 0.2\% up to 4\% in $H_2$ regions (Spitzer 1978). 
These are lower limits because large, massive dust grains do not have 
a detectable extinction. Direct satellite-based measurements give $r$-values 
up to 2\% (Frisch \etal 1999).  Little is yet known about the true 
dust-to-gas ratio in external galaxies, but some galaxies show large values.
E.g.\ M51 has a dust-to-gas ratio of 2\%. Taking the observed relation between 
the dust-to-gas ratio and the metallicity (Issa et al.\ 1990) into account, 
it is reasonable to assume larger dust mass fractions in the metal-enriched 
central regions of galaxies than on average.

In this paper we study the dynamical influence of a dust component
on circumnuclear gaseous disks. We restrict our analysis
to systems without a nuclear or a large scale bar and we do not
consider any formation or destruction processes of the dust. 
This investigation is done by 2D multi-component hydrodynamical simulations. 

\section{Method}

\subsection{Equations of motion}
\label{purehydrodynamics}

  We studied numerically the hydrodynamical equations for a 2-dimensional
single- or multi-component disk. Thus, we solved the continuity equation
\begin{equation}
    \pdert{\Sigma_{g,d}} +
         \nabla \cdot (\Sigma_{g,d} \vec{v}_{g,d}) = 0
\end{equation}
and the momentum equations for gas and dust. The momentum equations read
for gas
\begin{equation}
    \pdert{\vec{v}_g} + (\vec{v}_g \cdot \nabla) \vec{v}_g
        + \frac{\nabla P_g}{\Sigma_g}
        + \nabla (\Phi + \Phi_{\rm HBSD}) = S_g(\vec{v}_g) 
\end{equation}
\noindent
and for dust
\begin{equation}
    \pdert{\vec{v}_d} + (\vec{v}_d \cdot \nabla) \vec{v}_d
        + \nabla (\Phi + \Phi_{\rm HBSD}) = S_d(\vec{v}_d) \,\, .
\end{equation}
The components are
denoted by $g$ for the gaseous phase and by $d$ for the dust.
$\Sigma_{g,d}$ are the surface densities and $\vec{v}_{g,d}$ the
velocities. $P_g$ is the pressure of the gas which is given in the
2d-case as force per unit length.
$\Phi$ denotes the potential of the self-gravitating disk. 
It is derived from the Poisson equation 
\begin{equation}
  \Delta \Phi = 4 \pi G \Sigma(R,\varphi) \,\, \delta(z) 
              = 4 \pi G \left( \Sigma_g + \Sigma_d \right) \delta(z)\,\, .
\end{equation}
An external potential is added by a stationary contribution 
$\Phi_{\rm HBSD}$ related to the halo, the bulge and/or a stellar
disk component. These external potentials are chosen to match -- together
with the potential of the disk -- a given rotation curve. 

The main difference between the gaseous and the dust component is the
treatment of the dust as a pressureless phase. Therefore, if
gas and dust are in rotational equilibrium, there is a velocity
difference between both components which might give rise to a 
non-negligible frictional force depending on the cross-section for the gas-dust
interaction. This interaction is described by the source terms $S(\dots)$
on the RHS of the hydrodynamical equations. Since we do not
consider dust formation and destruction processes, the source terms
in the continuity equation vanish. However, frictional terms show up
in the equations of motion. The dust implementation will be described in the
next paragraph.

  The set of hydrodynamical equations is closed by a polytropic
equation of state
\begin{equation}
    P_g = K \Sigma_g^{\gamma_g} \,\,\, .
    \label{eqstate}
\end{equation}
For the gaseous phase we set the polytropic exponent to $\gamma_g=5/3$.
The constant $K$ is chosen to yield a given minimum Toomre parameter.

%
%
\subsection{Treatment of the dust component}
\label{dusttreatment}

The main difference between ''normal'' galactic disks and our dusty disks 
is that the cold component (dust) is not only coupled by gravity to the 
hotter phase (gas), but also by a frictional force between both components.
  This drag is taken into account by a source term in the
equations of motion following the general form suggested by Noh \etal 
(1991)
\begin{eqnarray}
   \vec{f} \equiv S_d(\vec{v}_d) & = & - A (\vec{v}_d - \vec{v}_g) \\
   S_g(\vec{v}_g) & = & - \frac{\Sigma_d}{\Sigma_g} \, \vec{f} \,\,.
   \label{eqsourcegas}
\end{eqnarray}
The second source term, Eq.\ (\ref{eqsourcegas}), follows from the 
requirement of momentum conservation. The physics of the friction is 
enclosed in the frictional timescale $A^{-1}$. 
Based on a microscopic view of the frictional
process, i.e.\ the momentum exchange between gas and dust particles
by collisions and the equipartition of momentum within the gaseous phase,
the timescales can be calculated. For the simulations shown in this paper
we assumed that the dust disk is thin compared to the gaseous disk.
In that case the frictional time scale is of the order of the dynamical
time scale given by the inverse circular frequency $\Omega^{-1}$
(Noh et al.\ 1991), i.e.\
\begin{equation}
    A = \Omega \,\,\, .
   \label{eqadynamics}
\end{equation}

The basic assumption is here that gas and dust establish very fast 
collisional equilibrium in the thin dust layer and that the gas momentum 
is then mixed vertically in a sound travelling time scale
(for further details see Theis \& Orlova 2004).

\subsection{Numerical implementation}
\label{numericalimplementation}

The nonlinear analysis implies the solution 
of the full set of hydrodynamical equations. For this purpose, we developed
a two-dimensional numerical code which is similar to the ZEUS-2D code 
by Stone \& Norman (1992). The hydrodynamical equations are 
discretized on a logarithmic Eulerian grid in polar coordinates (270$\times$270
grid cells). The different
terms are treated by operator splitting. Advection is performed by a
second order Van Leer advection scheme. 

\section{Results}

\subsection{Parameters of the initial gaseous disk}

  The characteristic values of our initial models are motivated by the nuclear 
region of M100 (NGC 4321). 
We adopted a total gas mass of $\natd{4.7}{8} \, \msun$ distributed 
exponentially with a scale length of 300 pc within a radial range of 
\mbox{$R_{\rm in} = 30 {\rm pc}$} to \mbox{$R_{\rm out} = 3 {\rm kpc}$}. 
The short disk scale length mimicks a central concentration of (molecular) 
gas observed in many galaxies. Small random perturbations are 
initially superimposed on the mass distribution by multiplying the
surface density in each cell with a factor $(1+A_R R)$. 
$R$ is a random number in the interval [-1,1] and $A_R$ 
a small amplitude of the order of $10^{-8}$.

The initial velocities are derived from rotational equilibrium, i.e.\ there is
no initial radial motion. The rotation curve is specified by 
\begin{equation}
   v_c(R) = v_\infty \cdot \frac{\displaystyle \frac{R}{R_{\rm flat}}}
    {\left[ 1 + \left(\displaystyle 
                 \frac{R}{R_{\rm flat}} \right)^{n_t} \right]^{1/n_t}}  \,\, .
   \label{eqvcirc}
\end{equation}
The velocity at infinity, $v_\infty$, was set to 178 km s$^{-1}$.
The transition parameter $n_t$ was selected to be 10 resulting in a
fairly sharp transition at the radius \mbox{$R_{\rm flat} = 100$ pc}. 
This rotation curve corresponds to a total dynamical mass 
$M_d(R) \sim v_c^2(R) R / G$ (including all components) 
of $\natd{7.3}{8} \msun$ within the central 100 pc.
In the region of rigid rotation the rotation period is about 
\mbox{$\natd{3.5}{6}$ yrs}. It increases outwards reaching
\mbox{$\natd{1.7}{7}$ yrs} at the half-mass radius of
the gaseous component at $R \approx 500$ pc.
The azimuthal velocity of the gaseous phase is calculated by the 
(frictionless) Jeans' equation. 

According to the chosen rotation curve, mass profile and equation of
state (polytropic with $\gamma_g=5/3$), the minimum value of the Toomre
parameter is reached at a galactocentric distance of about 440 pc.
The constant in the equation of state (for the gaseous phase) was selected
to yield a minimum Toomre parameter of $Q_{\rm min} = 1.54$.
This choice corresponds to sound speeds between 4 and 11 km s$^{-1}$ 
within the central kpc (the higher value is reached in the center). 

\begin{figure}[hp]
 \begin{center}
 \begin{tabular}{c}
  \psfig{file=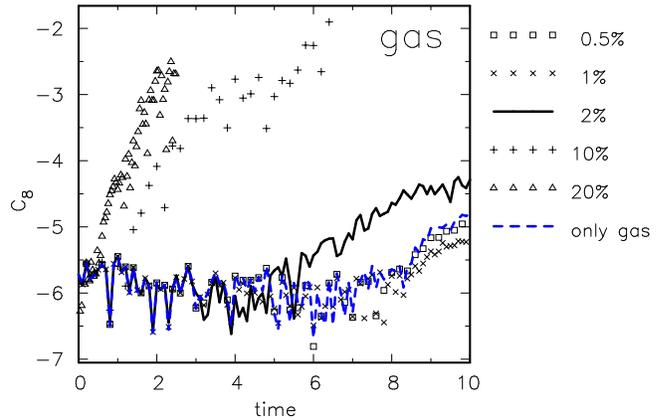,width=8.5cm,angle=270} \\
 \end{tabular}
   \caption{Temporal evolution of the logarithmic Fourier amplitudes 
    $C_8\equiv \log A_8$ of the 
    (dominant) $m=8$-mode of the dust component for different
    dust-to-gas mass fractions: 0.5\% (open boxes), 1\% (crosses),
    2\% (reference model, solid line), 10\% (plus), 20\% (triangle)
    and the purely gaseous model (dashed line).
    The time unit is $\natd{1.5}{7}$ yrs.}
   \label{globalmodes_dusttogas_dust_m8} 
 \end{center}
\end{figure}


\subsection{A dusty disk}

  We studied the influence of the dust-to-gas mass ratio $r$ by varying
$r\equiv M_d/M_g$ from 0.5\% to 20\%. 
  Fig.\ \ref{muchdust_surface_denspert_image} shows the 
surface density perturbations of gas and dust in the saturation stage
of a model with a large dust-to-gas ratio:
the irregular multi-armed structure is emphasized due to large
arm-interarm variations. Along the arms strong surface density
variations can be discerned. Some arms are interrupted by low density areas,
others are not smoothly curved, but show wiggles. Some arms seem
to merge with others. The dust distribution is highly correlated with
the gas distribution. However, the contrast between arm and interarm
regions is larger for the dust than for the gas. The surface
densities of the dust vary by about one order of magnitude, whereas
the contrast of the gas component is usually less than a factor of 2.
The structures formed in the dust are also thinner than those of the gas. 
Though there is a tight correlation between 
the positions of the maxima of the gas and dust phases, there is no 
correlation between their maximum amplitudes. The dust distribution is
characterized by a more cellular appearance compared to the spiral-like
morphology of the gas. 

It is interesting to note that the dust is often located at the
boundaries around peaks of the gas distribution, preferentially at the
inner boundary. Other dust peaks are in regions with no or only weak 
gaseous density enhancements. And, of course, there are dust peaks 
at the same locations as those of the gas. Thus, one expects large spatial 
variations of the dust-to-gas ratio.

In order to quantify the (global) stability of the disk we use the global 
Fourier amplitudes of each component
\begin{equation}
     A_m  \equiv {1 \over {M_{\rm disk}}}
     \left| \displaystyle \int_{0}^{2 \pi} \int_{R_{\rm in}}^{R_{\rm out}}
     \Sigma(r,\phi) r dr \, e^{-im\phi} d\phi \right| \,\,\,\,\,
     \mbox{\rm ($m>0$).}
     \label{eqamplitude}
\end{equation}
\noindent
$M_{\rm disk}$ is the mass of the disk for the component of interest
in the specified radial interval $[R_{\rm in},R_{\rm out}]$.
$\Sigma$ denotes the corresponding surface density. 
Comparing the modes from $m=1$ to $m=16$, the high-$m$-modes around 
$m \sim 8$ are dominant during the linear regime.
A comparison of the corresponding Fourier amplitudes with those 
of a purely gaseous model shows that the critical dust-to-gas ratio 
$r_c$ for dust becoming dynamically unimportant is about 1\% 
(Fig.\ \ref{globalmodes_dusttogas_dust_m8}). For larger amounts of dust the 
destabilization of the gaseous phase becomes much stronger. For 
$r=0.02$ the instability sets in about 50 Myrs
earlier than in dustless or low-dust-content models while the modes remain on
their initial value for a latency period of about 75 Myrs. 
Increasing $r$ to 10\% reduces the latency time to almost zero. 
The growth rates increase by a factor of 3-4 and the saturation level 
reached already after 30 Myrs is larger by at least one order of magnitude.

  A more detailed discussion on the method and the results as well as 
an extended parameter study can be found in our forthcoming paper 
(Theis \& Orlova 2004).

\section*{Summary}

   We studied the influence of a cold dust component on the evolution
of galactic gaseous disks by means of 2-dimensional hydrodynamical 
simulations. From the evolution of the Fourier amplitudes we found that the 
higher-order modes are the dominant unstable modes. Their growth is mainly 
restricted to the central kpc. An admixture of 2\% dust (relative to the gas 
mass) destabilizes the gaseous disk. The growth of instabilities in the dust 
component becomes non-linear after 100 Myr, followed by the gas 50 Myrs later. 

  The structures formed in both, gas and dust, are rather irregular and
multi-armed. This is a direct consequence of the superposition of many high-$m$
modes of similar amplitude. The dust component is characterized by a much 
stronger contrast between arm and inter-arm regions than the gas. The dust is 
spatially correlated with gas, but it does not exactly follow the gas. 
Peaks in the dust distribution are frequently found at the inner edges of 
peaks in the gas distribution. This results in a large
scatter of dust-to-gas ratios at different places. The dust develops also
thin filaments which sometimes connect the arms. Therefore, the dust
distribution has a more cellular appearance, whereas the gas develops
a multi-armed spiral morphology. 

  Below a dust-to-gas mass ratio of 1\% the dynamical influence of 
the dust on the gaseous disk becomes negligible. This critical value is 
close to the observed mean value in normal galaxies like the Milky Way. 
Since the dust-to-gas ratio scales linearily with metallicity, larger local
values of $r$, especially in the central galactic regions, seem to be
reasonable. Such values are also in agreement with local gas-to-dust
determinations. Since already a dust-to-gas ratio of 2\% significantly 
affects the evolution of the disk, even the observed small dust admixtures
are expected to have an impact on the dynamics of some galaxies 
(e.g.\ the dust-rich M51). For a 10\% admixture of dust the gaseous 
component is completely destabilized. The growth rates are enhanced by a 
factor of 3-4 with almost no latency phase. The saturation levels
reached after 30 Myr are substantially larger than in the
low-$r$ calculations. 

\section*{Acknowledgements}

The authors thank Vladimir Korchagin for stimulating discussions about disk 
stability. This work has been supported by the DFG under grants TH511/2-3 and 
RUS 436 RUS 17/65/02. C.T. is also grateful to the organizers of the GCD-V 
meeting for a very enjoyable conference as well as for financial support making
his participation possible. The simulations were performed at the computing 
center of the Christian-Albrechts-Universit\"at Kiel.

\section*{References}

\reference Athanassoula, E., 1992, MNRAS, 259, 345

\reference Binney, J., Tremaine, S., 1987, Galactic Dynamics,
                Princeton Univ.\ Press 

\reference Carollo, C.M., Stiavelli, M., de Zeeuw, P.T., Mack, J. 1997, 
               AJ, 114, 2366


\reference Elmegreen, B.G., Elmegreen, D.M., Eberwein, K.S. 2002, ApJ, 564, 234

\reference Englmaier, P., Shlosman I., 2000, ApJ, 528, 677

\reference Frisch, P.C., Dorschner, J.M., Geiss, J., \etal, 1999, ApJ, 525, 492

\reference Ikeuchi, S., Habe, A., Tanaka, Y.D., 1984, MNRAS, 207, 909

\reference Issa, M.R., MacLaren, I., Wolfendale, A.W. 1990, A\&A, 236, 237

\reference Jog, C.J., Solomon, P.M., 1984, ApJ, 276, 114

\reference Noh, H., Vishniac, E.T., Cochran, W., 1991, ApJ, 383, 372


\reference Regan, M.W., Mulchaey, J.S., 1999, AJ, 117, 2676

\reference Samland, M., Hensler, G., Theis, Ch., 1997, ApJ, 476, 544

\reference Spitzer, L., 1978, Physical Processes in the Interstellar Medium,
                 Wiley (New York)
                 
\reference Stone, J.M., Norman, M.L., 1992, ApJS, 80, 753

\reference Theis Ch., Burkert A., Hensler G., 1992, A\&A, 265, 465

\reference Theis Ch., Orlova, N., 2004, submitted to A\&A

\reference Wada, K., Koda, J., 2001, PASJ, 53, 1163

\end{document}